\documentclass[journal,comsoc]{IEEEtran}

\usepackage{amsmath}
\usepackage{url}
\usepackage{tabto}
\usepackage{xparse}
\usepackage{cite}
\usepackage{graphicx} 
\usepackage{makecell}
\usepackage{enumerate}
\usepackage[font=footnotesize,skip=0.1pt]{caption}
\usepackage{textcomp}
\usepackage{algorithm,algorithmic}
\usepackage{dingbat}
\usepackage{color}

\ExplSyntaxOn
\DeclareExpandableDocumentCommand{\eval}{m}{\int_eval:n {#1}}
\ExplSyntaxOff

\newcommand\tabsp{3}

\begin{document}
\title{Auto-Scaling Network Resources using Machine Learning to Improve QoS and Reduce Cost}

\author{\IEEEauthorblockN{Sabidur Rahman\IEEEauthorrefmark{1},
Tanjila Ahmed\IEEEauthorrefmark{1},
Minh Huynh\IEEEauthorrefmark{2}, 
Massimo Tornatore\IEEEauthorrefmark{1}\IEEEauthorrefmark{3}, and
Biswanath Mukherjee\IEEEauthorrefmark{1}}\\
\IEEEauthorblockA{\IEEEauthorrefmark{1}University of California, Davis, USA    \IEEEauthorrefmark{2}AT\&T Labs, USA    \IEEEauthorrefmark{3}Politecnico di Milano, Italy}\\
\IEEEauthorblockA{Email: \{krahman, tanahmed, mahuynh, mtornatore, bmukherjee\}@ucdavis.edu}}

\maketitle

\begin{abstract}
Virtualization of network functions (as virtual routers, virtual firewalls, etc.) enables network owners to efficiently respond to the increasing dynamicity of network services. Virtual Network Functions (VNFs) are easy to deploy, update, monitor, and manage. The number of VNF instances, similar to generic computing resources in cloud, can be easily scaled based on load. Hence, auto-scaling (of resources without human intervention) has been receiving attention. Prior studies on auto-scaling use measured network traffic load to dynamically react to traffic changes. In this study, we propose a proactive Machine Learning (ML) based approach to perform auto-scaling of VNFs in response to dynamic traffic changes. Our proposed ML classifier learns from past VNF scaling decisions and seasonal/spatial behavior of network traffic load to generate scaling decisions ahead of time. Compared to existing approaches for ML-based auto-scaling, our study explores how the properties (e.g., start-up time) of underlying virtualization technology impacts Quality of Service (QoS) and cost savings. We consider four different virtualization technologies: Xen and KVM, based on hypervisor virtualization, and Docker and LXC, based on container virtualization. Our results show promising accuracy of the ML classifier using real data collected from a private ISP. We report in-depth analysis of the learning process (learning-curve analysis), feature ranking (feature selection, Principal Component Analysis (PCA), etc.), impact of different sets of features, training time, and testing time. Our results show how the proposed methods improve QoS and reduce operational cost for network owners. We also demonstrate a practical use-case example (Software-Defined Wide Area Network (SD-WAN) with VNFs and backbone network) to show that our ML methods save significant cost for network service leasers.\footnote{\label{myfootnote}A short summarized version of this study was presented at the IEEE ICC 2018 conference in May 2018.}
\end{abstract}


\section{Introduction}
Network functions, such as those implemented in firewall, switch, router, Customer Premises Equipment (CPE), etc. have been traditionally deployed on proprietary hardware equipment, referred as “middleboxes”. This makes equipment upgrade, deployment of new features, and maintenance to be complex and time consuming for network owners. Virtualization of network functions can enable faster service deployment and flexible management~\cite{etsi}. Virtual Network Functions (VNFs) also allow us to use Commercial-Off-The-Shelf (COTS) hardware to replace costly vendor hardware.  VNFs are usually hosted inside cloud data centers (DCs), or in smaller metro data centers (e.g., Central Office Re-architected as a Datacenter (CORD)~\cite{cord}), or inside core network nodes.\\
\hspace*{\eval{\tabsp}mm}Auto-Scaling is traditionally used in cloud computing, where the amount of computational resources scales automatically based on load~\cite{amazon}. Auto-Scaling is an important mechanism for VNF management and orchestration; and it can reduce the operational cost for network owners (e.g., telco-cloud operators, DC operators). Also network leasers (e.g., mobile virtual network operators, enterprise customers) can benefit from flexible usage and pay-per-use pricing models enabled by auto-scaling.\\
\hspace*{\eval{\tabsp}mm}Traditional auto-scaling typically uses \emph{reactive} threshold-based approaches, that are simple to implement. But thresholds are difficult to choose and unresponsive to handle dynamic traffic. Recent studies suggest \emph{proactive} VNF scaling by combining traffic prediction and threshold-based methods. But using such methods for auto-scaling can lead to sub-optimal decisions. Hence, we propose a \emph{proactive} Machine Learning (ML) classifier to produce scaling decisions (instead of traffic prediction) ahead of time.\\
\hspace*{\eval{\tabsp}mm}Our proposal converts the VNF auto-scaling problem into a supervised ML classification problem. To train the supervised ML classifier, we use past VNF scaling decisions and measured network loads. The classification output is number of VNF instances required to serve future traffic without violating Quality of Service (QoS) requirements and deploying unnecessary VNF instances. We also provide in-depth analysis of the learning process (learning-curve analysis), feature ranking (feature selection, Principal Component Analysis (PCA), etc.), and impact of different sets of features, which has not been studied in the literature.\\
\hspace*{\eval{\tabsp}mm}Another contribution of this study is the analysis of the impact of underlying virtualization technology on auto-scaling performance. We compare four different virtualization technologies: Xen and KVM, based on hypervisor virtualization, and Docker and LXC, based on container virtualization. A virtual machine (VM) is an application/operating system (OS) environment that runs on host OS, but imitates dedicated hardware. In contrary, a container is a lightweight, portable, and stand-alone package of a software that can run on top of a host OS or host VM. Depending on the deployment scenario, critical properties of these VMs/containers such as start-up time can be significantly different.\\
\hspace*{\eval{\tabsp}mm}Our study compares accuracy of the proposed ML classifier using realistic network traffic load traces collected from an ISP. We discuss the impact of additional features and more training data on classification accuracy; and quantify how the proposed methods improve QoS and reduce operational cost for network owners and leasers over a practical use-case example (Software-Defined Wide Area Network (SD-WAN) with VNFs and backbone network).\\
\hspace*{\eval{\tabsp}mm}Significant contributions of our study are as follows:
\begin{enumerate}
\item We propose two ML-based methods which convert the auto-scaling decision to a ML classification problem, so that we can learn from the insights and temporal patterns hidden inside measured data from the network.
\item We explore accuracy of the ML classifiers using three performance metrics, for different categories of algorithms available in the ML suite WEKA~\cite{weka} and report results from seven top-performing ML algorithms.
\item We report in-depth analysis of the learning process (learning-curve analysis), feature ranking (feature selection, Principal Component Analysis (PCA), etc.), and impact of different sets of features. To the best of our knowledge, we are the first to report such detailed analysis for auto-scaling of network resources.
\item We also report the training (off-line model building) and testing (run-time decision making) time of the proposed ML classifiers, and explain how these parameters are crucial for practical implementation considerations.
\item Our results show how the proposed methods improve QoS, reduce operational cost for network owners, and reduce leasing cost for network service leasers. We also consider four different virtualization technology to compare their impact on VNF hosting.
\end{enumerate}
\hspace*{\eval{\tabsp}mm}The rest of this study is organized as follows. Section II reviews prior work on virtualization technologies and auto-scaling. Section III provides a formal problem statement for VNF auto-scaling. Section IV describes the proposed ML classifier and method. Section V discusses the performance of ML classifier, and illustrates numerical results on improvement of QoS and cost savings. Section VI concludes the study and indicates directions for future works.

\section{Background and Related Work}
VNFs can be deployed based on hypervisor virtualization or container virtualization. Compared to virtual machines (VMs), containers are lightweight, consume less CPU/memory/power resources, and have significantly less start-up time. Ref.~\cite{morabito} compares hypervisor-based virtualization (Xen, KVM, etc.) and container-based virtualization (Docker, LXC, etc.) in terms of power consumption. Containers require much less power, and this has a significant impact on operational cost. Choice of virtualization technology also impacts QoS. Ref.~\cite{piraghaj} discusses start-up time of different virtualization technologies, and shows how, after scaling decision, spawning a new VM/container takes longer/shorter time depending on the virtualization technology.\\
\hspace*{\eval{\tabsp}mm}Prior studies on auto-scaling can be classified in two groups: threshold-based vs. prediction-based (time series analysis, ML, etc.). Threshold-based approaches have been used by DC owners~\cite{amazon} for scaling computing resources. Static-threshold-based approaches~\cite{kana}\cite{murthy}\cite{hung} use predefined upper and lower thresholds for scaling, which is not practical in a dynamic demand scenario. Improvements have been proposed using dynamic threshold-based approaches~\cite{lorido}\cite{belo}\cite{lim}. As for prediction-based approaches, prior studies have used Auto-Regression (AR)~\cite{chandra}, Moving Average (MA)~\cite{mi}, and Auto-Regressive Moving Average (ARMA)~\cite{fang} to predict future workload for auto-scaling.\\
\hspace*{\eval{\tabsp}mm}Recent studies have explored scaling of VNFs in telco-cloud networks. Ref.~\cite{phung} proposes a deadline- and budget-constrained auto-scaling algorithm for 5G mobile networks. Here, VNFs are dynamically scaled based on number of current jobs in the system using a heuristic algorithm. Being reactive in nature, this method yields sub-optimal results in a network with fluctuating traffic. Ref.~\cite{tang} proposes a resource-efficient approach to auto-scale telco-cloud using reinforcement learning, and emphasizes that VNF auto-scaling is crucial for both QoS guarantee and cost management. Reinforcement learning runs without any knowledge of prior traffic pattern or scaling decisions. Hence, auto-scaling approaches in~\cite{phung}\cite{tang} do not benefit from the historic traffic data and scaling decisions.\\
\hspace*{\eval{\tabsp}mm}Ref.~\cite{rossem} covers auto-scaling from DevOps point of view. This study considers scaling data-plane resources as a function of traffic load; but also in this case, a reactive approach that does not benefit from historic data is used. Moreover, compared to~\cite{phung,tang,rossem}, our study explores the benefits of auto-scaling also from the leasers’ point of view.

\section{Problem Description}
The research problem of auto-scaling VNFs can be summarized as follows. Given:
\begin{itemize}
\item Set of VNF deployments (V), where each deployment \emph{v} $\epsilon$ V has one or more instances of VNFs serving network traffic. Each deployment \emph{v} has a minimum number of allowed VNF instances (\emph{v.min}) and a maximum number of allowed VNF instances (\emph{v.max}).
\item Backbone network topology G(D,E), where D is set of core network nodes and E is set of network links connecting the VNF deployments.  
\item Set of historic traffic load measurement data (H(\emph{v,t})) that requires processing from VNF deployment\emph{ v} at time \emph{t}.
\item Set of QoS requirement (Q(\emph{v})): For each VNF deployment \emph{v}, network traffic can tolerate delay, loss, etc.\footnote{\label{myfootnote}Prior studies~\cite{phung}\cite{tang} assume that this QoS requirement can be converted to a number of VNFs that can serve the traffic without violating QoS requirements. For example, one VNF instance can serve up to 1 Gbps line-rate traffic without violating QoS.}
\end{itemize}
\hspace*{\eval{\tabsp}mm}Our objective is to develop a method (see Fig.~\ref{example}.a) which dynamically scales the VNFs to minimize QoS violation and cost of operation/leasing. As we discuss in the next section, there is a trade-off between minimizing QoS violation and reducing cost. To reduce QoS degradation, the network owner needs to keep more resources running which adds to cost. In contrary, reducing cost requires less resource usage, which leads to QoS degradation, specially in case of bursty load. We propose two different methods focusing on optimizing each. The next section shows how we take this contradictory business-decision scenario and convert this problem to a ML classification problem. Our study focuses on the ``Auto-Scaling'' part of Fig.~\ref{example}.a and assumes that the ML classifier will be part of the orchestration and management modules of the network owner (similar to AT\&T's ECOMP~\cite{ecomp}). 

\section{Proposed ML Classifier}
\begin{figure}[!htb]
\centering
\includegraphics[width=3.0in]{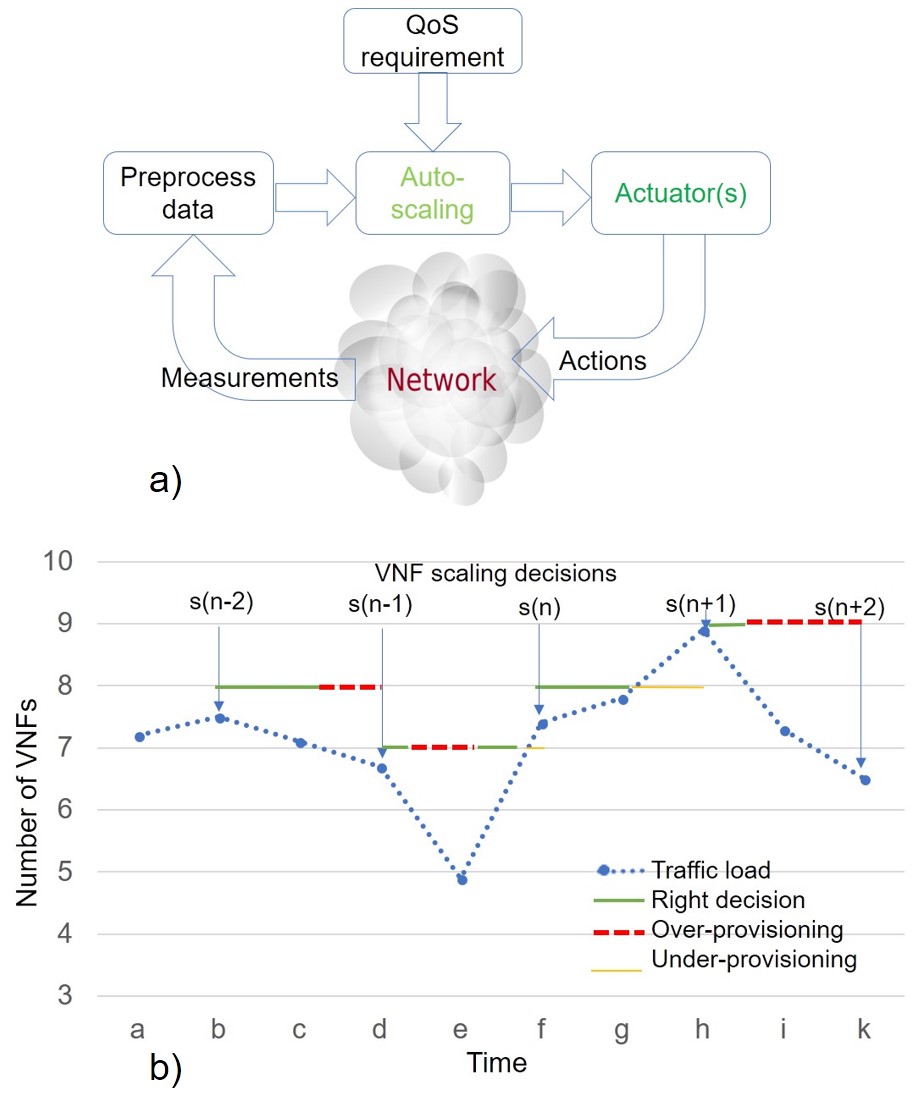}
\caption{Auto-Scaling overview: a) high-level view of auto-scaling decision life-cycle. b) example diagram to explain how traditional auto-scaling methods are easily vulnerable to over-provisioning and under-provisioning.}
\label{example}
\end{figure}

In machine learning, an \emph{instance} is a set of features/values representing a specific occurrence of the problem. For example, in our study, one \emph{feature} of the problem instance is \emph{time of day}. Another \emph{feature} is the value of the \emph{measured traffic load} at a {time of the day}. We can associate each instance (set of \emph{features}) of the problem to a \emph{class}, i.e., a classification decision. We convert the auto-scaling problem to a classification problem by training the classifier with a set of correctly-identified instances, called \emph{training set}. In the training phase, the ML classifier learns a mapping between the \emph{features} and \emph{classes}. After the training phase, a classifier can be tested using a set of instances, called \emph{test set}, which is not part of the \emph{training set}.\\
\hspace*{\eval{\tabsp}mm}The Time vs. ``Number of VNFs'' graph in Fig.~\ref{example}.b explains the input and output of the ML classifier. Fig.~\ref{example}.b shows different timestamps (\emph{a, b, c,} etc.) and auto-scaling decisions (\(s(n-1),s(n),s(n+1)\), etc.), where \(s(n)\) indicates the n-th auto-scaling decision. Let \emph{network traffic load} for given timestamp \emph{x} be $\lambda(x)$ and timestamp of auto-scaling step \emph{n} be given by $\tau(n)$. This figure also explains how auto-scaling decision can easily be prone to over-provisioning (red dashes, more VNFs than needed) and under-provisioning (yellow lines, less VNFs than required).

\subsection{Feature Selection (Input)}
\emph{Feature selection} is the first important step towards defining the ML classifier. All our \emph{features} are of numeric value. Referring to Fig.~\ref{example}, for auto-scaling decision $s(n+2)$, the proposed ML classifier maps recent traffic trends to a proactive scaling decision (\emph{class}). In this study, we convert the network traffic load measurements into the following features:
\begin{enumerate}
\item Day of month
\item Day of week
\item Weekday or weekend
\item Hour of day
\item Minute of hour
\item Timestamp of the decision $(k)$
\item Measured traffic at time \emph{k} $(\lambda(k))$
\item Traffic change from time \emph{j} to \emph{k} $(\lambda(k)-\lambda(j))$
\item Measured traffic at time \emph{j} $(\lambda(j))$
\item Traffic change from time \emph{i} to \emph{j} $(\lambda(j)-\lambda(i))$
\item Measured traffic at time \emph{i} $(\lambda(i))$
\item Traffic change from time \emph{h} to \emph{i} $(\lambda(i)-\lambda(h))$
\item Measured traffic at time \emph{h} $(\lambda(h))$
\item Traffic change from time \emph{g} to \emph{h} $(\lambda(h)-\lambda(g))$
\item Measured traffic at time \emph{g} $(\lambda(g))$
\item Traffic change from time \emph{f} to \emph{g} $(\lambda(g)-\lambda(f))$
\item Measured traffic at time \emph{f} $(\lambda(f))$
\item Traffic change from time \emph{e} to \emph{f} $(\lambda(f)-\lambda(e))$
\item Measured traffic at time \emph{e} $(\lambda(e))$
\item Traffic change from time \emph{d} to \emph{e} $(\lambda(e)-\lambda(d))$
\item Measured traffic at time \emph{d} $(\lambda(d))$
\item Traffic change from time \emph{c} to \emph{d} $(\lambda(d)-\lambda(c))$
\item Measured traffic at time \emph{c} $(\lambda(c))$
\item Traffic change from time \emph{b} to \emph{c} $(\lambda(c)-\lambda(b))$
\item Measured traffic at time \emph{b} $(\lambda(b))$
\item Traffic change from time \emph{a} to \emph{b} $(\lambda(b)-\lambda(a))$
\item Measured traffic at time \emph{a} $(\lambda(a))$
\end{enumerate}

\hspace*{\eval{\tabsp}mm}We consider up to 27 features containing temporal information of measured traffic load and traffic load change from recent past. Features 1-6 capture the temporal properties inside the data, and the rest of the features capture measured loads and how the loads change over time. We explain the impact of these features on the ML classifier using attribute selection algorithms, Principal Component Analysis (PCA), learning-curve analysis, etc. in later sections. In Section V, we also explain how we determine the number of features that provides the best accuracy for the classifier.

\subsection{Class Definition (Output)}
Next step is to define the output of the ML classifier, i.e., set of target \emph{classes} that the classifier tries to classify. In our study, \emph{class} depicts number of VNF instances, which is an integer value between \emph{v.min} and \emph{v.max}. To generate the labeled training set (\emph{instances} with known \emph{class} labels), we ensure that the scaling decision taken at step \emph{n} allocates enough resources to serve the traffic until the next decision-making step $n+1$.\\
\hspace*{\eval{\tabsp}mm}To define how we generate the \emph{class} value, we propose two different approaches:

\begin{enumerate}
\item QoS-prioritized ML (QML): In VNF auto-scaling, there is a trade-off between QoS and cost saving. We need to allocate more resources to guarantee QoS, but allocating more resources reduces cost saving. QML gives priority to QoS over cost saving. To guarantee QoS, auto-scaling decision at step $n$ (present) considers future traffic changes until next auto-scaling step $n+1$. QML generates the \emph{class} value as follows:
\begin{equation}
s(n)=max⁡(qos(\lambda(t))),\forall t \epsilon \{\tau(n),…,\tau(n+1)\}
\end{equation}
where \emph{t} is timestamp with traffic data between steps $n$ and $n+1$ (including $\tau(n)$ and $\tau(n+1)$). Function qos(.) takes the traffic load measured at time \emph{t} as input, and outputs the number of VNFs required to serve the measured traffic in line rate, without violating QoS.
\item Cost-prioritized ML (CML): In some cases, network owner/leaser may choose to ignore short-lived bursty traffic between steps $n$ and $n+1$ to save cost by avoiding over-provisioning of VNFs and accepting short-lived degradations. Auto-scaling decision for CML considers measured traffic load only at step $n$ (present) and at next auto-scaling step $n+1$ (future). CML generates the \emph{class} value as follows:
\begin{equation}
s(n)=max⁡(qos(\lambda(\tau(n))),qos(\lambda(\tau(n+1))))
\end{equation}
where $\tau(n)$ is the time at which step $n$ takes place and $\tau(n+1)$  is the time when step $n+1$ occurs.
\end{enumerate}

\subsection{Data Generation}
After defining the input and output of the classifier, the next task is dataset generation. For training-set and test-set generation, we assume that realistic traffic-load measurement data H($v,t$) is available at the network node where auto-scaling is performed. Table~\ref{knwonlabels} shows an example of training instance for QML and CML for scaling decision at steps $n$ and $n+1$ (Fig.~\ref{example}) where $f,g,h,i,$ etc. are time values.
\begin{table}[!htb]
\caption{Instances with Known Labels.}
\label{knwonlabels}
\centering
\begin{tabular}{|c|c|c|c|c|c|c|c|}
\hline
Case & \makecell{Fea.\\1} & \makecell{Fea.\\2} & \makecell{Fea.\\3} & \makecell{Fea.\\4} & ... & \makecell{QML\\Class} & \makecell{CML\\Class}\\
\hline
1 & g & $\lambda(g)$ & $\makecell{\lambda(g)-\\ \lambda(f)}$ & $\lambda(f)$ & & $qos(\lambda(h))$ & $\makecell{qos(\lambda(\tau\\(n)))}$\\
\hline
2 & i & $\lambda(i)$ & $\makecell{\lambda(i)-\\ \lambda(g)}$ & $\lambda(g)$ & & $qos(\lambda(h))$ & $\makecell{qos(\lambda(\tau\\(n+2)))}$\\
\hline
\end{tabular}
\end{table}

\subsection{Machine Learning Algorithms}
Selecting the right ML algorithm is the next task towards training the classifier. We explore different categories of algorithms in the ML suite WEKA~\cite{weka}, including decision-tree-based algorithms, rule-based algorithms, artificial neural networks, and Bayesian-network-based algorithms. Below is a brief introduction to the algorithms preforming well in our study:

\begin{enumerate}[(a)]
\item Random Tree: Random tree is the decision-tree implementation of WEKA. ``Random'' refers to a decision tree built on a random subset of columns. Decision trees classify instances by sorting them based on feature values. Each node in a decision tree is a feature, and each branch represents a value that the node uses to classify instances.
\item J48: \emph{C4.5} algorithm, proposed by Ross Quinlan is one of the most well-known algorithms to build decision trees~\cite{supervised}. J48 is WEKA's implementation of C4.5.
\item REPTree: Reduced Error Pruning (REP) tree is another improved version of decision-tree algorithm. REPTree benefits from information gain and minimization of error arising from variance.
\item Random Forest: Random forest improves the tree classifier by averaging/voting between several tree classifiers. Instead of building multiple trees on the same data, random forest adds randomness by building each tree on slightly-different rows and randomly-selected subset of columns.
\item Decision Table: Similar to tree-based algorithms, rule-based algorithms such as decision table try to infer decisions by learning the ``rule'' hidden inside the training instances. Each decision corresponds to a variable or relationship whose possible values cover the decisions.
\item Multi-Layer Perceptron (MLP): MLP is a class of feed-forward artificial neural networks, which are powerful classification and learning algorithms.
\item Bayesian Network (BayesNet): Bayesian network is a probabilistic model that represents a set of random variables and their conditional dependencies.
\end{enumerate}

\subsection{Performance Evaluation}
A test dataset is used to evaluate the performance of the trained classifier. Given a trained ML classifier and a test set, the test outcome is divided into four groups: i) True Positive (TP): positive instances correctly classified; ii) False Positive (FP): negative samples incorrectly classified as positive; iii) True Negative (TN): negative samples correctly classified; iv) False Negative (FN): positive samples wrongly classified as negative.\\
\hspace*{\eval{\tabsp}mm}We consider three different performance metrics:

\begin{enumerate}[(a)]
\item Precision (\%): Precision corresponds to the fraction of predicted positives which are in fact positive. Precision is a strong indication of accuracy for the ML classifier. Precision is given by percentage of: $TP/(TP+FP)$.
\item False Positive (\%): FP is an important indication of ML classifiers as lower FP indicates less classification mistakes.
\item ROC area: Receiver Operating Characteristic (ROC) curve is a graphical plot in which true positive rate $(TP/((TP+FN))$ is plotted as function of the false positive rate $(FP/(FP+TN))$. ROC area is a robust metric for ML classifier performance evaluation.
\end{enumerate}

\subsection{Conversion of Number of VNFs to Backbone Network Capacity}
\hspace*{\eval{\tabsp}mm}In Section V.H, we explore leasing cost of a SD-WAN use-case with VNFs and backbone network. We use the following equation to convert the output from ML classifier (number of VNFs required) to the required amount of bandwidth (C) over the transport network:
\begin{equation}
C= s(n)*Q'                                                          
\end{equation}
where $s(n)$ is auto-scaling decision at step $n$, and $Q'$ is the conversion unit for line-rate traffic processed by each VNF (e.g., one VNF can process upto 1 Gbps traffic).

\subsection{Operational Cost: Electricity Consumption Model}
Usually, VNFs, similar to virtual machines (VMs), are deployed inside a server which resides inside a rack. Rack power consumption depends on: number of physical servers hosted per rack, power consumption of individual servers, and utilization of those servers. In prior studies, a linear model is used to calculate the server power consumption:
\begin{equation}
P_{s} = P_{idle} + P_{peak} * u
\end{equation}
where,\\
Server utilization, $u$: current server load over max server load.\\
$P_{idle}$: Server idle power (when server does not have any load).\\
$P_{peak}$: Server peak power (when the server is at full load).\\
\hspace*{\eval{\tabsp}mm}We follow the model in~\cite{rahman} for server power consumption. This model is used to obtain results in Section V.G.

\subsection{Leasing Cost Model}
We propose this leasing cost model to show the cost minimization from the leaser’s point of view in Section V.H. The \emph{pay-per-use} model enabled by our approach is shown as:
\begin{equation}
C_{l} = C_{v} * v * \alpha  + C_{n} * b * \alpha + C_{q} * \beta                 
\end{equation}
where,\\
$C_{l}$: Total payment the leaser pays for the service.\\
$C_{v}$: Cost per unit of VM per second.\\
$C_{n}$: Cost per unit of bandwidth (Gbps) per second.\\
$C_{q}$: Cost per second due to degraded QoS. The revenue that the leaser looses if the service does not maintain QoS.\\
$v$: Required number of instances of VNF.\\
$b$: Required bandwidth (Gbps).\\
$\alpha$: Service usage duration (seconds).\\
$\beta$: Duration of service suffering from degraded QoS (seconds).\\
\section{Illustrative Numerical Examples}
This section shows the performance of the proposed ML classifier for auto-scaling. Our results show how the proposed approaches improve QoS and reduce cost for both network owners and leasers.
\subsection{Experimental Setup}
We consider that VNFs are hosted inside virtualized instances on top of physical servers, similar to generic VMs in DC deployments~\cite{piraghaj}. We assume that the maximum traffic load that requires processing at such a deployment is equivalent to 10 Gbps, and each VNF can handle maximum 1 Gbps traffic without degrading QoS. The acceptable range for number of VNFs is minimum (\emph{v.min}) one and maximum (\emph{v.max}) 10. For VNF auto-scaling, we consider ``horizontal'' scaling, i.e., removing or adding new VNF instances every time we scale in/out. When load increases, the network owner deploys additional VNF instances to satisfy QoS requirement. In case of decreasing load, the network owner removes some VNF instances to save operational cost (e.g., electricity usage).\\
\hspace*{\eval{\tabsp}mm}We use realistic traffic load traces from~\cite{oli}. Traffic load data (in bits) was collected at every five-minute intervals over a 1.5-month period from a private ISP and on a transatlantic link~\cite{traffic}. We use the traffic data to generate ``features'' and ``classes'' for training and testing both QML and CML. As traffic load traces are at every 5-minute interval, we assume auto-scaling decisions are made every 10 minutes. However, our proposed methods in Section IV are generic enough to work for other interval granularities.\\
\hspace*{\eval{\tabsp}mm}The ML algorithm settings used in WEKA~\cite{weka} follow:
\begin{itemize}
\item Random Forest: batch size = 100, number of iterations = 100.
\item Random Tree: batch size = 100, minimum variance proportion = 0.001.
\item J48: batch size = 100, confidence factor = 0.25.
\item REP Tree: batch size = 100, minimum variance proportion = 0.001.
\item Decision Table: batch size = 100, search method: bestfirst (hill climbing algorithm with backtracking).
\item MLP: batch size = 100, hidden layers = (number of attributes + number of classes) / 2, learning rate = 0.3.
\item BayesNet: batch size = 100, search algorithm = K2 (hill climbing algorithm).
\end{itemize}

\subsection{Auto-Scaling Accuracy of Proposed ML Classifiers}
\begin{figure}[!htb]
\centering
\includegraphics[width=3.0in]{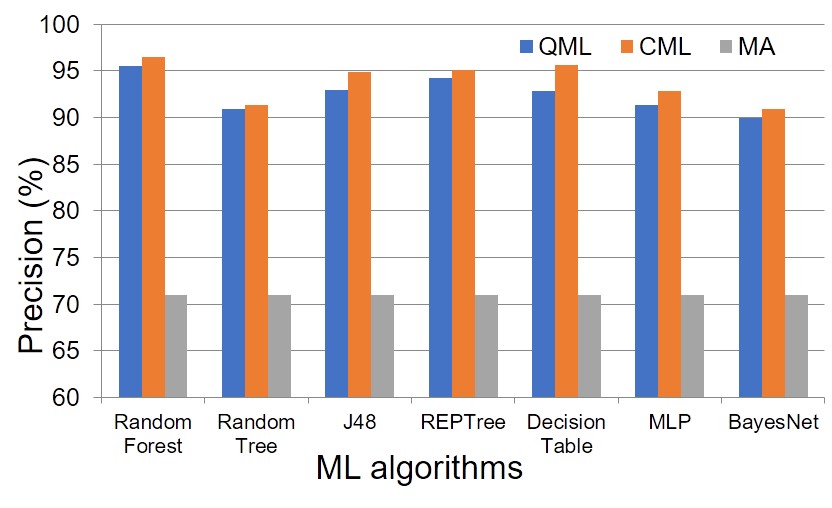}
\caption{Auto-Scaling accuracy (precision (\%)) of proposed ML classifiers.}
\label{precision}
\end{figure}

Here, we consider the three performance metrics discussed in Section IV.E to explain the accuracy of the proposed methods. In Fig.~\ref{precision}, we compare the auto-scaling accuracy (precision) of the proposed ML classifier. We compare QML (QoS-prioritized) and CML (cost-prioritized) approaches with prior study ``MA''~\cite{mi}. Prior study ``MA'' proposed a time-series-based prediction approach using Moving Average (MA) for proactive auto-scaling. For Fig.~\ref{precision}, we use 40 days of data ($40*144=5760$ instances) for training and two days of data ($2*144=288$ instances) for testing. We assume that the ML model is retrained with new data every two days. First, Fig.~\ref{precision} shows that the ML classifier takes auto-scaling decisions with higher accuracy for both QML (95.5\%) and CML (96.5\%) compared to MA (71\%) for the same test data. Second, Fig.~\ref{precision} shows that, among different ML algorithms used to train the classifier, ``Random Forest'' has higher precision for both QML and CML approaches. Difference in prediction accuracy is due to different ``class'' generation results (see Eqn. (1)) than CML (Eqn. (2)).

\begin{figure}[!htb]
\centering
\includegraphics[width=3.0in]{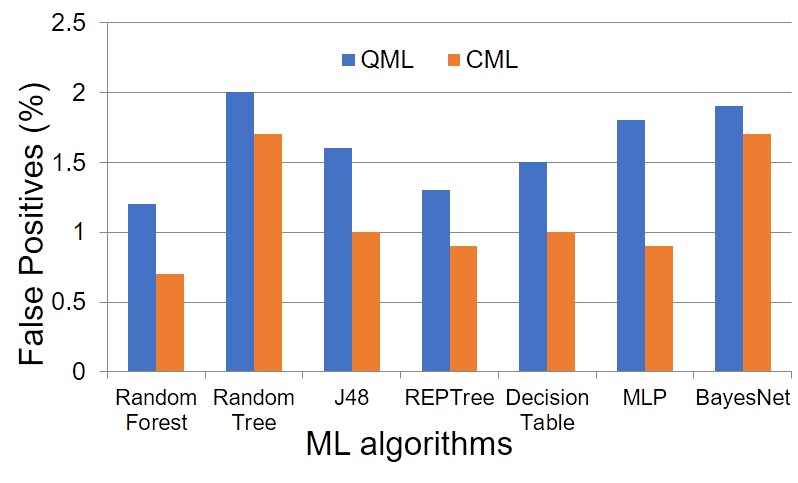}
\caption{Auto-Scaling accuracy (false positives (\%)) of proposed ML classifiers.}
\label{fprate}
\end{figure}

False Positives are important metric for ML classifiers. If a ML classifier generates too many false positives, in most scenarios, the classifier will not be considered as a recommended one. Fig.~\ref{fprate} reports the False Positives (lower is better) for QML and CML with different ML algorithms. ``Random Forest'' shows lowest FP with 1.2\% for QML and 0.7\% for CML. Results in Fig.~\ref{fprate} are generated with the same setup as in Fig.~\ref{precision}.

\begin{figure}[!htb]
\centering
\includegraphics[width=3.0in]{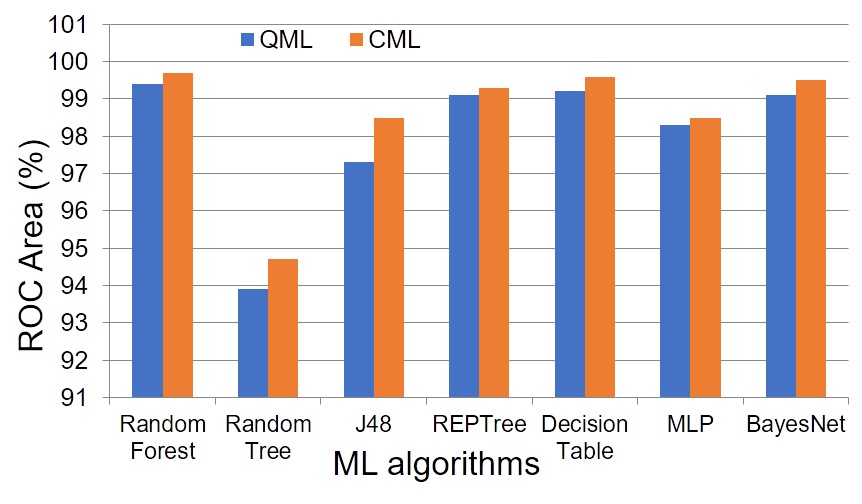}
\caption{Auto-Scaling accuracy (ROC Area (\%)) of proposed ML classifiers.}
\label{rocarea}
\end{figure}

ROC Area is another important and robust metric for ML classifiers. ROC Area considers the performance of the ML classifier in complete range of true positives and false positives, and then reports the overall performance of the classier. Fig. \ref{rocarea} shows the ROC Area (\%) for QML and CML (higher is better). Again, Random Forest shows highest ROC Area with 99.4\% for QML and 99.7\% for CML. Results in Fig.~\ref{rocarea} are generated with the same setup as in Fig.~\ref{precision}.

\hspace*{\eval{\tabsp}mm}Decision-tree-based algorithms perform better for the scaling decision compared to neural-network-based and Bayesian-network-based algorithms. Among decision-tree-based algorithms, ``Random Forest'' leads with highest precision (96.5\%), highest ROC Area (99.7\%), and lowest false positives (0.7\%). Clearly, the pattern of the data and feature set favor decision-tree-based algorithms to learn better than other ML techniques. But ``Random Forest'' further improves the performance by averaging/voting between several tree-classifier instances. For the rest of the numerical evaluations, we use the results from ``Random Forest'' to explore learning curves, QoS degradation, and leasing cost.

\subsection{Learning Curve Analysis: Impact of 'Number of Features' and 'Training Dataset Size' on Auto-Scaling Accuracy}
\begin{figure}[!htb]
\centering
\includegraphics[width=3.0in]{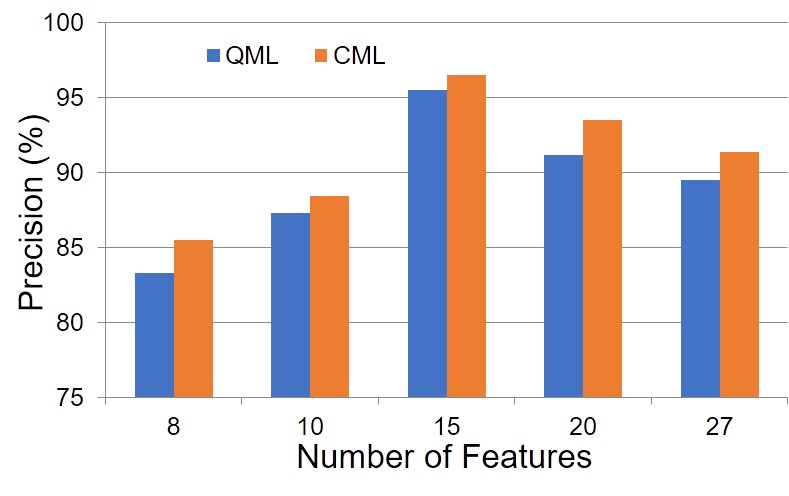}
\caption{Number of features vs. auto-scaling accuracy in precision (\%).}
\label{numfeatures}
\end{figure}

\begin{figure}[!htb]
\centering
\includegraphics[width=3.0in]{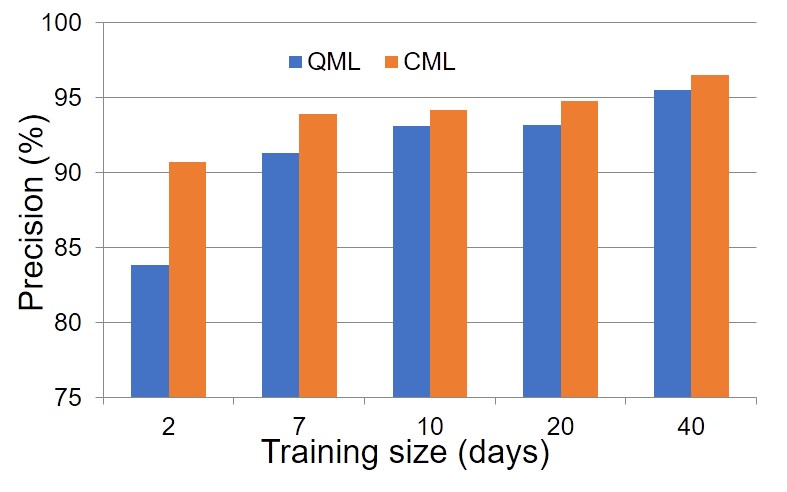}
\caption{Training data size vs. auto-scaling accuracy in precision (\%).}
\label{trainingsize}
\end{figure}

Figs.~\ref{numfeatures}-\ref{trainingsize} provide auto-scaling learning curve of the proposed ML classifier. Learning-curve analysis helps us to understand the following two important aspects of training a classifier: ``How many features generate best results?'' and ``Does more data help or not?''

\hspace*{\eval{\tabsp}mm}Fig.~\ref{numfeatures} shows the impact of number of features in the accuracy of auto-scaling decisions made by the ML classifier. For example, number of features ``8'' means we are using only the first eight features from Section IV.A. As shown in Fig.~\ref{numfeatures}, the accuracy of the ML classifier increases with the number of features. But, after the number of features goes higher than ``15'', the accuracy decreases. This means that, if we keep adding more features by moving away from the time of scaling decision, the additional \emph{features} impact the accuracy negatively. We discuss the impact of features in more detail in the next subsection.

\hspace*{\eval{\tabsp}mm}Fig.~\ref{trainingsize} shows the impact of training dataset size on the auto-scaling accuracy (precision) of the ML classifier. The general intuition is that, with more data, the ML classifier should perform better. We observe that \emph{7 days of training data} has significant performance improvement over \emph{2 days of training data}. One explanation of this phenomenon is that \emph{7 days of training data} offers insights from seasonal pattern and periodicity of the load during the whole week (compared to 2 days).  Then, \emph{10 days of training data} improves the ML model further, but, \emph{20 days of training data} does not have much additional learning points. Then, \emph{40 days of training data} introduces the monthly pattern and improves the precision significantly more than 20 days. For rest of the study, we consider 15 features and 40 days of training data.

\subsection{More Insights: Feature Ranks and Impact of Different Features on Auto-Scaling Accuracy}
\hspace*{\eval{\tabsp}mm}Important questions regarding ML classification are: ``Can we identify the dominant features from the feature list?'' ``Which features are impacting more towards classification accuracy?'' ``Can we explain how different combination of features are impacting the accuracy?'' In this subsection, we explore different methods to answer these questions.\\
\hspace*{\eval{\tabsp}mm}First, we use attribute (feature) selection algorithm \emph{InfoGainAttributeEval}~\cite{wekainfogain} from WEKA which evaluates the worth of a feature by measuring the information gain with respect to the classification. After ranking the first 15 features, Features 7, 9, 11, 13, and 15 are ranked 1 through 5, in that order. This observation gives two important insights: a) ``Measured loads'' are the most important features that contributed to accurate classification; and b) ``Measured loads'' closer to the decision time have more significant impact on the classification.\\
\hspace*{\eval{\tabsp}mm}We have confirmed this observation using Principal Component Analysis (PCA), a statistical procedure (often used with feature-ranking methods) to find correlated features and their impact on classification. In PCA, the first principal component has the largest possible variance which accounts for much of the variability in the data. Our PCA reports a combination of features 7, 9, 11, 13, and 15 as the first principal component, conveying similar takeaway as \emph{InfoGainAttributeEval}.\\
\hspace*{\eval{\tabsp}mm}Feature 2 (day of week), feature 1 (day of month), and feature 3 (weekday or weekend) are ranked 6th, 7th and 8th, respectively. As expected, these three features carry information related to the temporal variation of load, so they are ranked highly in feature ranks. Rest of the features are ranked in the following sequence: 14, 12, 6, 4, 8, 10, and 5.\\
\hspace*{\eval{\tabsp}mm}Table~\ref{featureimpact} shows impact of different features on auto-scaling decision accuracy. We compare the precision of the algorithms with different combination of features such as ``Measured Load'' (features 7, 9, 11, 13, and 15), DoM (Day of Month), DoW (Day of week), W (Weekday or Weekend), HoD (Hour of Day), etc. As discussed earlier, only ``Measured load'' feature set shows very high precision. Then, in second row, ``Measured Load'' with the rest of the temporal features improves accuracy up to 96.2\%. Only the temporal features (row 3) show significant accuracy of 83\%. One interesting observation is that, if we pick a single temporal feature with ``Measured Load'' features, the performance degrades. This means temporal features can help to improve decisions only when they work together to provide the complete seasonal variations and patterns.
\begin{table}[!htb]
\caption{Impact of Different Features.}
\label{featureimpact}
\centering
\begin{tabular}{|c|c|c|c|c|c|}
\hline
Measured Load & DoM & DoW & W & HoD & CML\\
\hline
\checkmark & & & & & 96.1\\
\hline
\checkmark & \checkmark & \checkmark & \checkmark & \checkmark & 96.2\\
\hline
 & \checkmark & \checkmark & \checkmark & \checkmark & 83\\
\hline
\checkmark & \checkmark &  &  & & 95.8\\
\hline
\checkmark & & & \checkmark & & 95.5\\
\hline
\end{tabular}
\end{table}

\subsection{Training and Testing Time of ML Classifiers}
\hspace*{\eval{\tabsp}mm}Our study assumes the ML classifier will run in real-time to provide auto-scaling decisions. We also assume that the model will be retrained every two days with updated data. Hence, it is important to report the training (off-line model building) and testing (run-time decision making) times of the proposed ML classifiers. Table~\ref{time} shows the training time (5760 instances, 40 days of data) and testing time (288 instances, two days of data) for different ML algorithms. WEKA reports the times in seconds upto two digits after decimal point. This means the zeros reported in the table takes milliseconds or less time to make 288 auto-scaling decisions, which is very promising for real-time deployment for our proposed method. Also, training times are few seconds or less, which supports retraining the model every two days.\\
\hspace*{\eval{\tabsp}mm}To compare the algorithms, ``MLP'' (neural network) takes longest (16.69s) and ``Random Tree'' takes shortest (0.05s) to train the models. In run-time, we see many sub-millisecond algorithms such as ``Decision Table''. In the contrary, ``Random Forest'' takes 0.02 seconds. This is an important decision-making point. For example, in a special case, if the network owner is willing to accept slight loss of accuracy (Decision Table 95.6\% vs. Random Tree 96.5\%) to obtain faster decisions, ``Decision Table'' can be a better choice than ``Random Forest''. Such practical consideration related to ML-based solutions is a strong motivation for our study.

\begin{table}[!htb]
\caption{Training and Testing Time of Auto-Scaling ML Classifiers.}
\label{time}
\centering
\begin{tabular}{|c|c|c|}
\hline
ML Algorithm & Training Time (s) & Testing Time (s)\\
\hline
Random Forest & 1.77 & 0.02\\
\hline
Random Tree & 0.05 & 0\\
\hline
J48 & 0.28 & 0.02\\
\hline
REPTree & 0.15 & 0\\
\hline
Decision Table & 0.7 & 0\\
\hline
MLP & 16.69 & 0\\
\hline
BayesNet & 0.17 & 0.01\\
\hline
\end{tabular}
\end{table}

\subsection{Impact of Auto-Scaling on QoS}
\begin{figure}[!htb]
\centering
\includegraphics[width=3.0in]{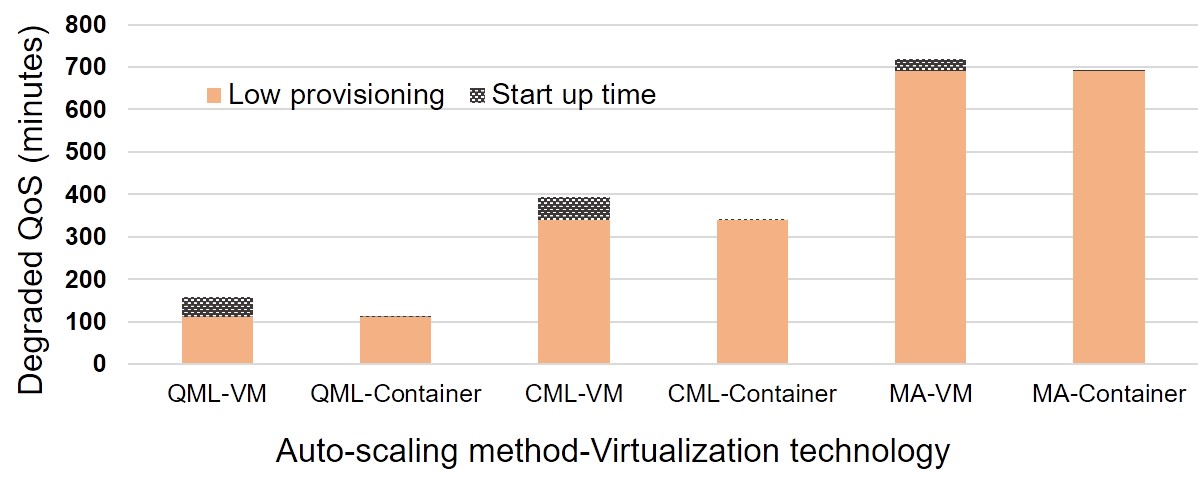}
\caption{Impact of auto-scaling methods and virtualization technologies on QoS (minutes).}
\label{qos}
\end{figure}

To explore the impact of auto-scaling (using the proposed ML classifier) and virtualization technology on QoS-guarantee, Fig.~\ref{qos} compares QML, CML, and MA in terms of ``Degraded QoS''. We define ``Degraded QoS'' as the cumulative time (minutes) a VNF deployment spends in QoS violation. We explore two reasons behind ``Degraded QoS'': i) Low provisioning: The proactive method failed to allocate enough resources to accommodate future traffic; and ii) Start-up time: Even if the auto-scaling method decides to turn on more VNFs, spawning a new VM takes time. The impact on start-up time is expected to be different for different virtualization technologies. Hypervisor-based VMs and containers are compared in Fig.~\ref{qos}. The results are taken from the two-day test output from ML classifier using ``Random Forest'' algorithm.\\

\hspace*{\eval{\tabsp}mm}Fig.~\ref{qos} shows that QML spends significantly less time in ``Degraded QoS'' compared to CML and MA. We also observe that choice of virtualization technology can make a difference in terms of QoS as containers have much faster start time (0.4 seconds) than hypervisor-based VMs (100 seconds)~\cite{piraghaj}, and this helps containers to improve QoS significantly.

\subsection{Impact of Auto-Scaling on Operational Cost}

\begin{figure}[!htb]
\centering
\includegraphics[width=3.0in]{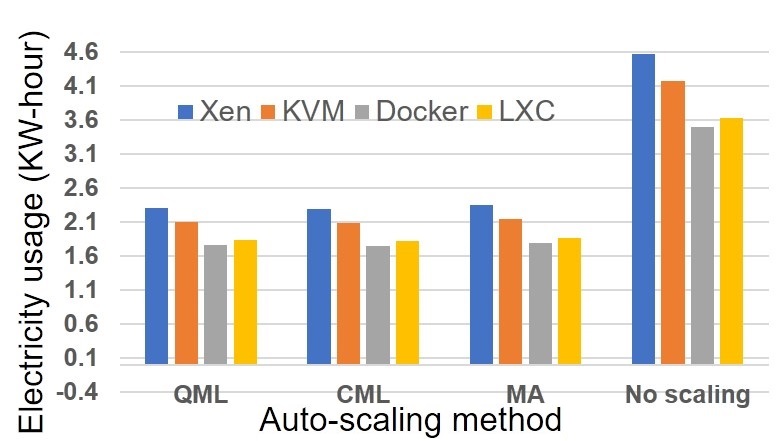}
\caption{Impact of auto-scaling methods and virtualization technologies on operational cost (electricity usage).}
\label{electricity}
\end{figure}

\hspace*{\eval{\tabsp}mm}Fig.~\ref{electricity} explores the electricity usage of different virtualization technologies and scaling methods. Power consumption values for VMs (XEN, KVM) and containers (Docker, LXC) are taken from~\cite{morabito}. We find that both QML and CML consume less electricity than MA. But compared to CML, QML keeps more VNFs running to avoid degraded QoS, resulting in higher electricity usage. Docker containers shows least usage of electricity compared to other technologies (Xen, KVM, LXC, etc.). Electricity usage, degradation of services, and other operational activities can be converted into operational cost values as well using appropriate cost models~\cite{rahman}.

\subsection{Impact of Auto-Scaling on Leasing Cost}
\begin{figure}[!htb]
\centering
\includegraphics[width=3.0in]{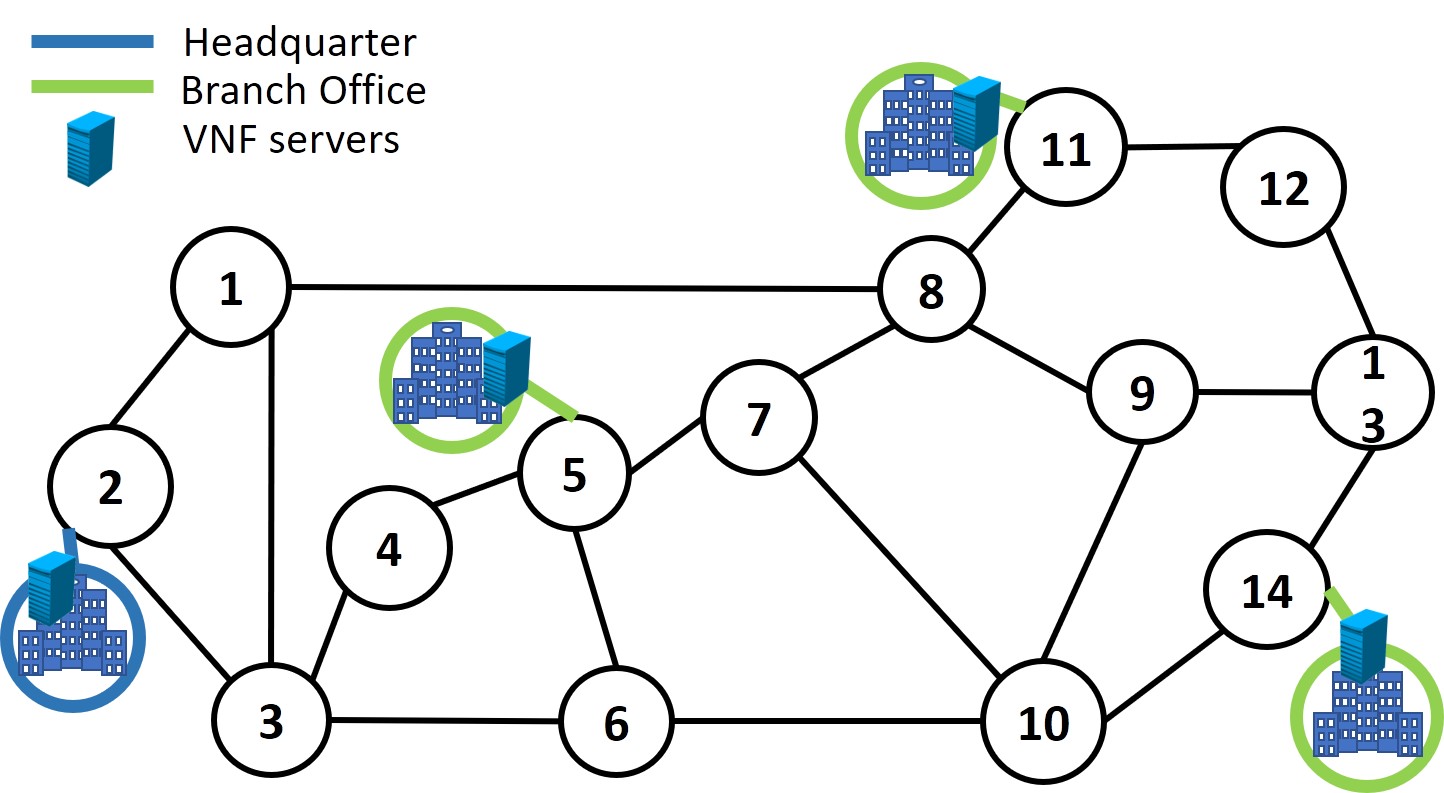}
\caption{Use-case scenario for SD-WAN with backbone network.}
\label{topo}
\end{figure}

We assume now an SD-WAN use-case, where a leaser leases VNFs and connectivity over a backbone transport network. Fig.~\ref{topo} shows the backbone network topology and an enterprise network with one headquarter and three branch offices connected by the network. For each of the four offices, VNFs are deployed on servers. For our study, we assume that a service chain ~\cite{gupta} with three VNF services (namely, firewalls, routers, and private branch exchanges (PBXs)) are deployed in all sites. We also assume that each office experiences traffic load same as the two-day test dataset. We use Eqn. (3) to convert the output (auto-scaling decisions) from ML classifier to allocate the required connectivity over the network. For example, if ML classifier outputs the number of VNFs required to serve the traffic to be two, and each VNF can serve up to 1 Gbps line-rate traffic, then the network requires 2 Gbps capacity to carry the traffic without violating QoS.\\

\begin{figure}[!htb]
\centering
\includegraphics[width=3.0in]{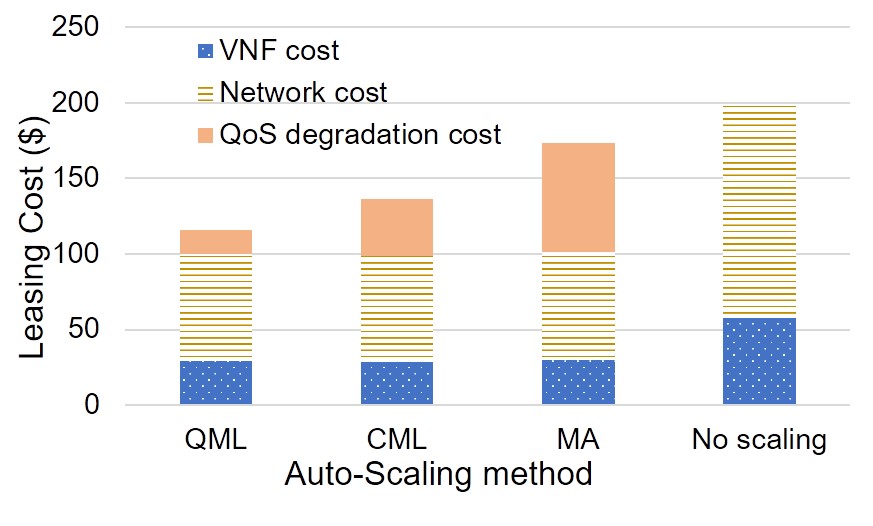}
\caption{Impact of auto-scaling methods on leasing cost (\$).}
\setlength{\belowcaptionskip}{-10pt}
\label{leasing}
\end{figure}

\hspace*{\eval{\tabsp}mm}Fig.~\ref{leasing} compares the cost of VNFs, network, and service degradation for different auto-scaling approaches. For VNF, we consider the leasing cost of \$0.01/second/VM from Google cloud~\cite{cloud}. For backbone network, we consider leasing cost \$70/Gbs/month from Google fiber~\cite{fiber}. For degraded service, we assume the enterprise loses \$1 for every 10 minutes of degraded service. Leasing cost shown in Fig.~\ref{leasing} is derived from testing on two days of data using ``Random Forest'' algorithm.\\

\hspace*{\eval{\tabsp}mm}Fig.~\ref{leasing} shows that QML ensures the lowest leasing cost. However, by considering just VNF cost and network cost, CML ensures lower cost than QML and MA. This means enterprises which are interested to reduce cost by sacrificing QoS will benefit from our CML method. On the other hand, enterprises which cannot afford degradation of service (i.e., QoS violation is costly) will benefit from our QML method.

\section{Conclusion}
Our study proposed a Machine Learning method to perform VNF auto-scaling. The ML classifier learns from historic VNF auto-scaling decisions and shows promising accuracy rate (96.5\%). Illustrative results show the impact of number of features and training-data size on the proposed ML classifier. We also studied the impact of start-up time of four different virtualization technologies on QoS. We explored a practical SD-WAN use-case with a backbone network showing that our proposed method yields lower leasing cost for network leasers compared to prior works. Future studies should consider exploring detailed analysis of operational and leasing costs for more such use-cases.  In our study, we explored only horizontal scaling of VNFs. Vertical scaling (i.e., adding/removing CPU/memory resources to the same virtual instance) for VNFs is an important direction yet to be explored. As network services are often deployed as “service chains”, future studies should explore auto-scaling methods for such scenarios.

\end{document}